\documentclass{elsart}
\usepackage{graphicx}
\usepackage{amsmath}
\usepackage{amsfonts}
\usepackage{amssymb}

\begin{document}
\begin{frontmatter}
\title
{On the electromagnetic field of a charged collapsing spherical shell in general relativity}
\author[salerno,icra]{Christian Cherubini}\ead{cherubini@icra.it},
\author[icra]{Remo Ruffini}\ead{ruffini@icra.it},
\author[icra]{Luca Vitagliano}\ead{vitagliano@icra.it}
\address
[salerno]{Dipartimento di Fisica ``E. R. Caianiello", Universit\`a di Salerno, 84081 Baronissi (SA), Italy}
\address
[icra]{International Centre for Relativistic Astrophysics, Department of Physics,
Rome University ``La Sapienza", P.le Aldo Moro 5, 00185 Rome, Italy}
\begin{abstract}
A new exact solution of the Einstein-Maxwell equations for the gravitational collapse of a shell of matter
in an already formed black hole is given. Both the shell and the black hole are endowed with
electromagnetic structure and are assumed spherically symmetric. Implications for current research
are outlined.
\end{abstract}
\end{frontmatter}

It is well known that the role of exact solutions has been fundamental in the
development of general relativity. We present here a new exact solution for a
charged shell of matter collapsing into a black hole endowed with
electromagnetic structure, for short an EMBH. For simplicity we consider the
case of zero angular momentum and spherical symmetry. This problem is relevant
for its own sake as an addition to the existing family of interesting exact
solutions and also represents some progress in understanding the role of the
formation of the horizon and of the irreducible mass introduced in
\cite{CR71}, see e.g. \cite{RV02}. It is also essential to a rigorous
treatment of the vacuum polarization processes occurring during the formation
of an EMBH, see e.g. \cite{RVX02}. Both of these issues are becoming relevant
to explaining gamma ray bursts, see e.g.
\cite{RBCFX01a,RBCFX01b,RBCFX01c,RBCFX01d} and references therein.

W. Israel and V. de La Cruz \cite{I66,IdlC67} showed that the problem of a
collapsing charged shell can be reduced to a set of ordinary differential
equations. We reconsider here the following relativistic system: a spherical
shell of electrically charged dust which is moving radially in the
Reissner-Nordstr\"{o}m background of an already formed nonrotating EMBH of
mass $M_{1}$ and charge $Q_{1}$, with $Q_{1}\leq M_{1}$. . The
Einstein-Maxwell equations with a charged spherical dust as source are
\begin{equation}
G_{\mu\nu}=8\pi\left[  T_{\mu\nu}^{\left(  \mathrm{d}\right)  }+T_{\mu\nu
}^{\left(  \mathrm{em}\right)  }\right]  ,\quad\nabla_{\mu}F^{\nu\mu}=4\pi
j^{\nu},\quad\nabla_{\lbrack\mu}F_{\nu\rho]}=0,
\end{equation}
where
\begin{equation}
T_{\mu\nu}^{\left(  \mathrm{d}\right)  }=\varepsilon u_{\mu}u_{\nu},\quad
T_{\mu\nu}^{\left(  \mathrm{em}\right)  }=\tfrac{1}{4\pi}\left(  F_{\mu}%
{}^{\rho}F_{\rho\nu}-\tfrac{1}{4}g_{\mu\nu}F^{\rho\sigma}F_{\rho\sigma
}\right)  ,\quad j^{\mu}=\sigma u^{\mu}.
\end{equation}
Here $T_{\mu\nu}^{\left(  \mathrm{d}\right)  }$, $T_{\mu\nu}^{\left(
\mathrm{em}\right)  }$ and $j^{\mu}$ are respectively the energy-momentum
tensor of the dust, the energy-momentum tensor of the electromagnetic field
$F_{\mu\nu}$ and the charge $4-$current. The mass and charge density in the
comoving frame are given by $\varepsilon$, $\sigma$ and $u^{a}$ is the
$4$-velocity of the dust. In spherical-polar coordinates the line element is
\begin{equation}
ds^{2}\equiv g_{\mu\nu}dx^{\mu}dx^{\nu}=-e^{\nu\left(  r,t\right)  }%
dt^{2}+e^{\lambda\left(  r,t\right)  }dr^{2}+r^{2}d\Omega^{2},
\end{equation}
where $d\Omega^{2}=d\theta^{2}+\sin^{2}\theta d\phi^{2}$.

We describe the shell by using the $4$-dimensional Dirac distribution
$\delta^{\left(  4\right)  }$ normalized as
\begin{equation}
\int\delta^{\left(  4\right)  }\left(  x,x^{\prime}\right)  \sqrt{-g}d^{4}x=1
\end{equation}
where $g=\det\left\|  g_{\mu\nu}\right\|  $. We then have
\begin{align}
\varepsilon\left(  x\right)   &  =M_{0}\int\delta^{\left(  4\right)  }\left(
x,x_{0}\right)  r^{2}d\tau d\Omega,\label{eq1a}\\
\sigma\left(  x\right)   &  =Q_{0}\int\delta^{\left(  4\right)  }\left(
x,x_{0}\right)  r^{2}d\tau d\Omega. \label{eq2a}%
\end{align}
$M_{0}$ and $Q_{0}$ respectively are the rest mass and the charge of the shell
and $\tau$ is the proper time along the world surface $S:$ $x_{0}=x_{0}\left(
\tau,\Omega\right)  $ of the shell. $S$ divides the space-time into two
regions: an internal one $\mathcal{M}_{-}$ and an external one $\mathcal{M}%
_{+}$. As we will see in a forthcoming work \cite{RV02} for the description of
the collapse we can choose either $\mathcal{M}_{-}$ or $\mathcal{M}_{+}$. The
two description, clearly equivalent, will be relevant for the physical
interpretation of the solutions.

Introducing the orthonormal tetrad
\begin{equation}
{\boldsymbol\omega}_{\pm}^{\left(  0\right)  }=f_{\pm}^{1/2}dt,\quad
{\boldsymbol\omega}_{\pm}^{\left(  1\right)  }=f_{\pm}^{-1/2}dr,\quad
{\boldsymbol\omega}^{\left(  2\right)  }=rd\theta,\quad{\boldsymbol\omega
}^{\left(  3\right)  }=r\sin\theta d\phi;\quad
\end{equation}
we obtain the tetrad components of the electric field%
\begin{equation}
{\boldsymbol{\mathcal E}}=\mathcal{E\ }{\boldsymbol\omega}^{\left(  1\right)
}=\left\{
\begin{array}
[c]{l}%
\frac{Q}{r^{2}}\ {\boldsymbol\omega}_{+}^{\left(  1\right)  }\qquad
\text{outside the shell}\\
\frac{Q_{1}}{r^{2}}\ {\boldsymbol\omega}_{-}^{\left(  1\right)  }%
\qquad\text{inside the shell}%
\end{array}
\right.  , \label{E3}%
\end{equation}
where $Q=Q_{0}+Q_{1}$ is the total charge of the system. From the $G_{tt}$
Einstein equation
\begin{equation}
ds^{2}=\left\{
\begin{array}
[c]{l}%
-f_{+}dt_{+}^{2}+f_{+}^{-1}dr^{2}+r^{2}d\Omega^{2}\qquad\text{outside the
shell}\\
-f_{-}dt_{-}^{2}+f_{-}^{-1}dr^{2}+r^{2}d\Omega^{2}\qquad\text{inside the
shell}%
\end{array}
\right.  , \label{E0}%
\end{equation}
where $f_{+}=1-\tfrac{2M}{r}+\tfrac{Q^{2}}{r^{2}}$, $f_{-}=1-\tfrac{2M_{1}}%
{r}+\tfrac{Q_{1}^{2}}{r^{2}}$ and $t_{-}$ and $t_{+}$ are the
Schwarzschild-like time coordinates in $\mathcal{M}_{-}$ and $\mathcal{M}_{+}$
respectively. Here $M$ is the total mass-energy of the system formed by the
shell and the EMBH, measured by an observer at rest at infinity.

Indicating by $R$ the Schwarzschild radius of the shell and by $T_{\pm}$ its
time coordinate, from the $G_{tr}$ Einstein equation we have
\begin{equation}
\tfrac{M_{0}}{2}\left[  f_{+}\left(  R\right)  \tfrac{dT_{+}}{d\tau}%
+f_{-}\left(  R\right)  \tfrac{dT_{-}}{d\tau}\right]  =M-M_{1}-\tfrac
{Q_{0}^{2}}{2R}-\tfrac{Q_{1}Q_{0}}{R}. \label{eq3a}%
\end{equation}
The remaining Einstein equations are identically satisfied. From (\ref{eq3a})
and the normalization condition $u_{\mu}u^{\mu}=-1$ we find
\begin{align}
\left(  \tfrac{dR}{d\tau}\right)  ^{2}  &  =\tfrac{1}{M_{0}^{2}}\left(
M-M_{1}+\tfrac{M_{0}^{2}}{2R}-\tfrac{Q_{0}^{2}}{2R}-\tfrac{Q_{1}Q_{0}}%
{R}\right)  ^{2}-f_{-}\left(  R\right) \nonumber\\
&  =\tfrac{1}{M_{0}^{2}}\left(  M-M_{1}-\tfrac{M_{0}^{2}}{2R}-\tfrac{Q_{0}%
^{2}}{2R}-\tfrac{Q_{1}Q_{0}}{R}\right)  ^{2}-f_{+}\left(  R\right)
,\label{EQUY}\\
\tfrac{dT_{\pm}}{d\tau}  &  =\tfrac{1}{M_{0}f_{\pm}\left(  R\right)  }\left(
M-M_{1}\mp\tfrac{M_{0}^{2}}{2R}-\tfrac{Q_{0}^{2}}{2R}-\tfrac{Q_{1}Q_{0}}%
{R}\right)  . \label{EQUYa}%
\end{align}

We now define $r_{\pm}\equiv M\pm\sqrt{M^{2}-Q^{2}}$: when $Q<M$, $r_{\pm}$
are real and they correspond to the horizons of the new black hole formed by
the gravitational collapse of the shell. We similarly introduce the horizons
$r_{\pm}^{1}=M_{1}\pm\sqrt{M_{1}^{2}-Q_{1}^{2}}$ of the already formed EMBH.
From (\ref{eq3a}) we have
\begin{equation}
M-M_{1}-\tfrac{Q_{0}^{2}}{2R}-\tfrac{Q_{1}Q_{0}}{R}>0 \label{Constraint}%
\end{equation}
holds for $R>r_{+}$ if $Q<M$ and for $R>r_{+}^{1}$ if $Q>M$ since in these
cases the left hand side of (\ref{eq3a}) is clearly positive. Equations
(\ref{EQUY}) and (\ref{EQUYa}) (together with (\ref{E0}), (\ref{E3}))
completely describe a 5-parameter ($M$, $Q$, $M_{1}$, $Q_{1}$, $M_{0}$) family
of solutions of the Einstein-Maxwell equations.

For astrophysical applications \cite{RVX02} the trajectory of the shell
$R=R\left(  T_{+}\right)  $ is obtained as a function of the time coordinate
$T_{+}$ relative to the space-time region $\mathcal{M}_{+}$. In the following
we drop the $+$ index from $T_{+}$. From (\ref{EQUY}) and (\ref{EQUYa}) we
have
\begin{equation}
\tfrac{dR}{dT}=\tfrac{dR}{d\tau}\tfrac{d\tau}{dT}=\pm\tfrac{F}{\Omega}%
\sqrt{\Omega^{2}-F}, \label{EQUAISRDLC}%
\end{equation}
where
\begin{equation}
F\equiv f_{+}\left(  R\right)  =1-\tfrac{2M}{R}+\tfrac{Q^{2}}{R^{2}}%
,\quad\Omega\equiv\Gamma-\tfrac{M_{0}^{2}+Q^{2}-Q_{1}^{2}}{2M_{0}R}%
,\quad\Gamma\equiv\tfrac{M-M_{1}}{M_{0}}.
\end{equation}
Since we are interested in an imploding shell, only the minus sign case in
(\ref{EQUAISRDLC}) will be studied. We can give the following physical
interpretation of $\Gamma$. If $M-M_{1}\geq M_{0}$, $\Gamma$ coincides with
the Lorentz $\gamma$ factor of the imploding shell at infinity; from
(\ref{EQUAISRDLC}) it satisfies
\begin{equation}
\Gamma=\tfrac{1}{\sqrt{1-\left(  \frac{dR}{dT}\right)  _{R=\infty}^{2}}}\geq1.
\end{equation}
When $M-M_{1}<M_{0}$ then there is a \emph{turning point} $R^{\ast}$, defined
by $\left.  \tfrac{dR}{dT}\right|  _{R=R^{\ast}}=0$. In this case $\Gamma$
coincides with the ``effective potential'' at $R^{\ast}$ :
\begin{equation}
\Gamma=\sqrt{f_{-}\left(  R^{\ast}\right)  }+M_{0}^{-1}\left(  -\tfrac
{M_{0}^{2}}{2R^{\ast}}+\tfrac{Q_{0}^{2}}{2R^{\ast}}+\tfrac{Q_{1}Q_{0}}%
{R^{\ast}}\right)  \leq1.
\end{equation}

The solution of the differential equation (\ref{EQUAISRDLC}) is given by:
\begin{equation}
\int dT=-\int\tfrac{\Omega}{F\sqrt{\Omega^{2}-F}}dR. \label{GRYD}%
\end{equation}
The functional form of the integral (\ref{GRYD}) crucially depends on the
degree of the polynomial $P\left(  R\right)  =R^{2}\left(  \Omega
^{2}-F\right)  $, which is generically two, but in special cases has lower
values. We therefore distinguish the following cases:

\begin{enumerate}
\item {\boldmath$M=M_{0}+M_{1}$}; {\boldmath$Q_{1}=M_{1}$}; {\boldmath
$Q=M$}: $P\left(  R\right)  $ is equal to $0$, we simply have
\begin{equation}
R(T)=\mathrm{{const}.}%
\end{equation}

\item {\boldmath$M=M_{0}+M_{1}$}; {\boldmath$M^{2}-Q^{2}=M_{1}^{2}-Q_{1}^{2}$%
}; {\boldmath$Q\neq M$}: $P\left(  R\right)  $ is a constant, we have
\begin{equation}
T=\mathrm{const}+\tfrac{1}{2\sqrt{M^{2}-Q^{2}}}\left[  \left(  R+2\right)
R+r_{+}^{2}\log\left(  \tfrac{R-r_{+}}{M}\right)  +r_{-}^{2}\log\left(
\tfrac{R-r_{-}}{M}\right)  \right]  . \label{CASO1}%
\end{equation}

\item {\boldmath$M=M_{0}+M_{1}$}; {\boldmath$M^{2}-Q^{2}\neq M_{1}^{2}%
-Q_{1}^{2}$}: $P\left(  R\right)  $ is a first order polynomial and
\begin{align}
T  &  =\mathrm{const}+2R\sqrt{\Omega^{2}-F}\left[  \tfrac{M_{0}R}{3\left(
M^{2}-Q^{2}-M_{1}^{2}+Q_{1}^{2}\right)  }\right. \nonumber\\
&  \left.  +\tfrac{\left(  M_{0}^{2}+Q^{2}-Q_{1}^{2}\right)  ^{2}%
-9MM_{0}\left(  M_{0}^{2}+Q^{2}-Q_{1}^{2}\right)  +12M^{2}M_{0}^{2}%
+2Q^{2}M_{0}^{2}}{3\left(  M^{2}-Q^{2}-M_{1}^{2}+Q_{1}^{2}\right)  ^{2}%
}\right] \nonumber\\
&  -\tfrac{1}{\sqrt{M^{2}-Q^{2}}}\left[  r_{+}^{2}\mathrm{arctanh}\left(
\tfrac{R}{r_{+}}\tfrac{\sqrt{\Omega^{2}-F}}{\Omega_{+}}\right)  -r_{-}%
^{2}\mathrm{arctanh}\left(  \tfrac{R}{r_{-}}\tfrac{\sqrt{\Omega^{2}-F}}%
{\Omega_{-}}\right)  \right]  , \label{CASO2}%
\end{align}

where $\Omega_{\pm}\equiv\Omega\left(  r_{\pm}\right)  $.

\item {\boldmath$M\neq M_{0}+M_{1}$}: $P\left(  R\right)  $ is a second order
polynomial and
\begin{align}
T &  =\mathrm{const}-\tfrac{1}{2\sqrt{M^{2}-Q^{2}}}\left\{  \tfrac
{2\Gamma\sqrt{M^{2}-Q^{2}}}{\Gamma^{2}-1}R\sqrt{\Omega^{2}-F}\right.
\nonumber\\
&  +r_{+}^{2}\log\left[  \tfrac{R\sqrt{\Omega^{2}-F}}{R-r_{+}}+\tfrac
{R^{2}\left(  \Omega^{2}-F\right)  +r_{+}^{2}\Omega_{+}^{2}-\left(  \Gamma
^{2}-1\right)  \left(  R-r_{+}\right)  ^{2}}{2\left(  R-r_{+}\right)
R\sqrt{\Omega^{2}-F}}\right]  \nonumber\\
&  -r_{-}^{2}\log\left[  \tfrac{R\sqrt{\Omega^{2}-F}}{R-r_{-}}+\tfrac
{R^{2}\left(  \Omega^{2}-F\right)  +r_{-}^{2}\Omega_{-}^{2}-\left(  \Gamma
^{2}-1\right)  \left(  R-r_{-}\right)  ^{2}}{2\left(  R-r_{-}\right)
R\sqrt{\Omega^{2}-F}}\right]  \nonumber\\
&  -\tfrac{\left[  2MM_{0}\left(  2\Gamma^{3}-3\Gamma\right)  +M_{0}^{2}%
+Q^{2}-Q_{1}^{2}\right]  \sqrt{M^{2}-Q^{2}}}{M_{0}\left(  \Gamma^{2}-1\right)
^{3/2}}\log\left[  \tfrac{R}{M}\sqrt{\Omega^{2}-F}\right.  \nonumber\\
&  \left.  \left.  +\tfrac{2M_{0}\left(  \Gamma^{2}-1\right)  R-\left(
M_{0}^{2}+Q^{2}-Q_{1}^{2}\right)  \Gamma+2M_{0}M}{2M_{0}M\sqrt{\Gamma^{2}-1}%
}\right]  \right\}  .\label{CASO3}%
\end{align}
\end{enumerate}

In the case of a shell falling in a flat background ($M_{1}=Q_{1}=0$) it is of
particular interest to study the \emph{turning points} $R^{\ast}$ of the shell
trajectory. In this case equation (\ref{EQUY}) reduces to
\begin{equation}
\left(  \tfrac{dR}{d\tau}\right)  ^{2}=\tfrac{1}{M_{0}^{2}}\left(
M+\tfrac{M_{0}^{2}}{2R}-\tfrac{Q^{2}}{2R}\right)  ^{2}-1. \label{EQUY2}%
\end{equation}
Case $(2)$ has no counterpart in this new regime and eq. (\ref{Constraint})
constrains the possible solutions to only the following cases:

\begin{enumerate}
\item {\boldmath$M=M_{0}$}; {\boldmath$Q=M_{0}$. }$R=R\left(  0\right)  $ constantly.

\item {\boldmath$M=M_{0}$}; {\boldmath$Q<M_{0}$. }There are no turning points,
the shell starts at rest at infinity and collapses until a
Reissner-Nordstr\"{o}m black-hole is formed with horizons at $R=r_{\pm}\equiv
M\pm\sqrt{M^{2}-Q^{2}}$ and the singularity in $R=0$.

\item {\boldmath$M\neq M_{0}$. }There is one turning point $R^{\ast}$.

\begin{enumerate}
\item {\boldmath$M<M_{0}$}, then necessarily is {\boldmath$Q<M_{0}$}.
Positivity of rhs of (\ref{EQUY2}) requires $R\leq R^{\ast}$, where $R^{\ast
}=\frac{1}{2}\frac{Q^{2}-M_{0}^{2}}{M-M_{0}}$ is the unique turning point.
Then the shell starts from $R^{\ast}$ and collapses until the singularity at
$R=0$ is reached.

\item {\boldmath$M>M_{0}$}. The shell has finite radial velocity at infinity.

\begin{enumerate}
\item {\boldmath$Q\leq M_{0}$}. The dynamics are qualitatively analogous to
case (2).

\item {\boldmath$Q>M_{0}$}. Positivity of the rhs of (\ref{EQUY2}) and
(\ref{Constraint}) requires that $R\geq$ $R^{\ast}$, where $R^{\ast}=\frac
{1}{2}\frac{Q^{2}-M_{0}^{2}}{M-M_{0}}$. The shell starts from infinity and
bounces at $R=R^{\ast}$, reversing its motion.
\end{enumerate}
\end{enumerate}
\end{enumerate}

In this regime the analytic forms of the solutions are given by eqs.
(\ref{CASO2}) and (\ref{CASO3}), simply setting $M_{1}=Q_{1}=0$.

\begin{figure}[th]
\begin{center}
\includegraphics[width=12cm]{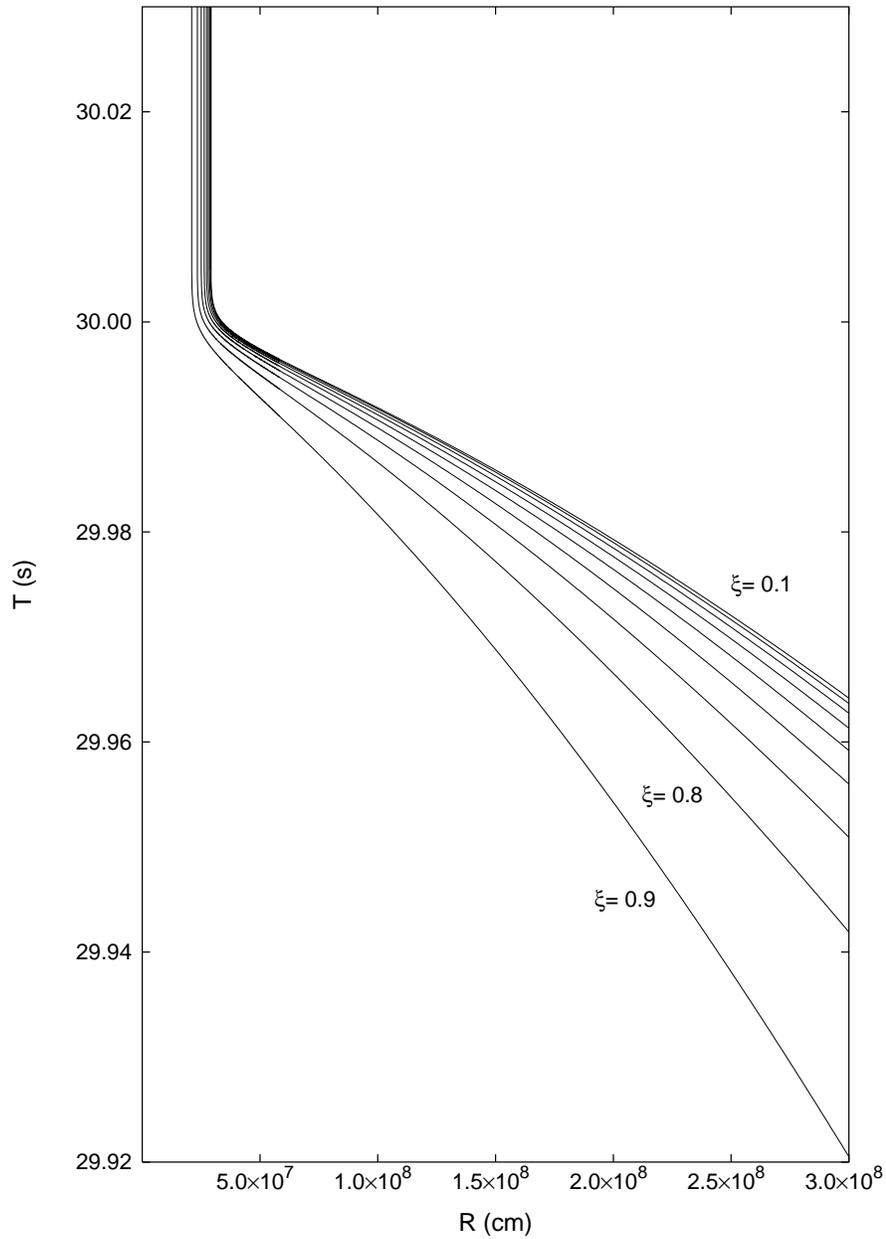}
\end{center}
\caption{Collapse curves in the plane $(T,R)$ for $M=20M_{\odot}$ and for
different values of the parameter $\xi$. The asymptotic behavior is the clear
manifestation of general relativistic effects as the horizon of the EMBH is
approached}%
\label{fig1}%
\end{figure}

\begin{figure}[th]
\begin{center}
\includegraphics[width=12cm]{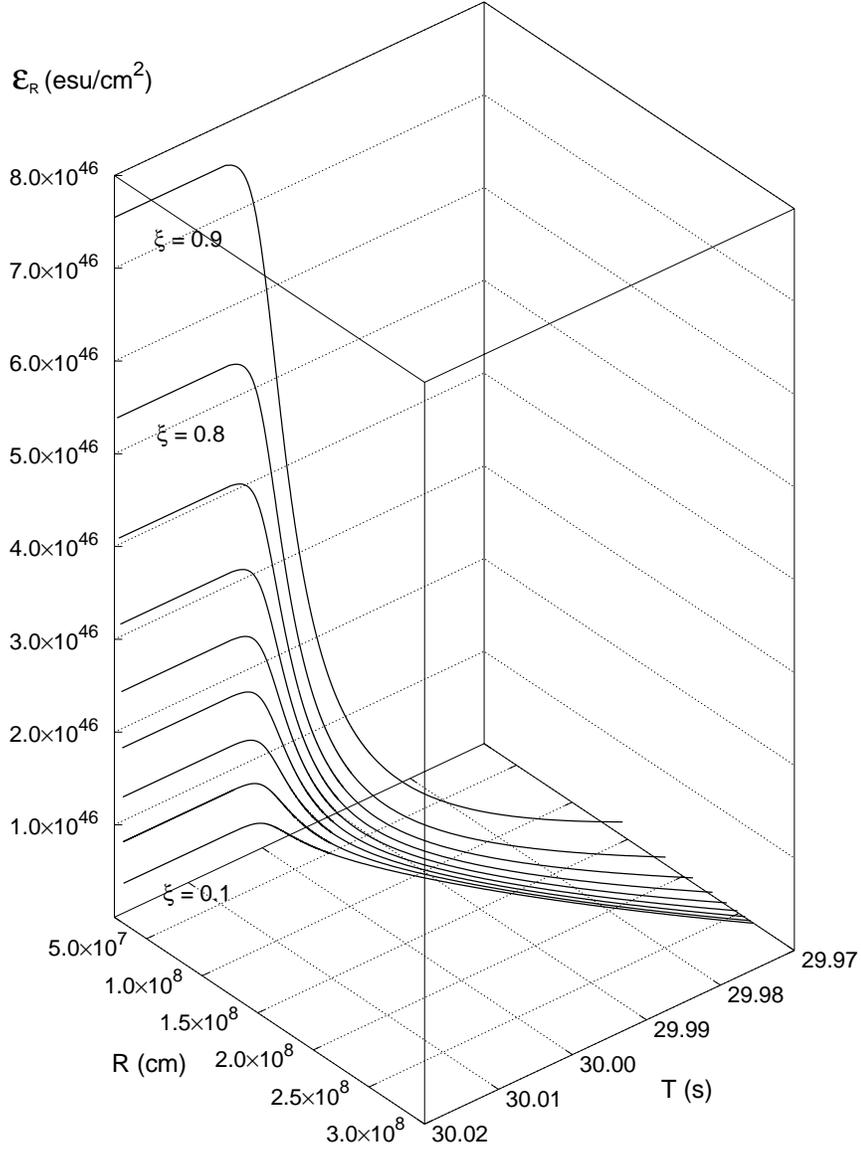}
\end{center}
\caption{Electric field behaviour at the surface of the shell for $M=20
M_{\odot}$ and for different values of the parameter $\xi$. The asymptotic
behavior is the clear manifestation of general relativistic effects as the
horizon of the EMBH is approached}%
\label{fig2}%
\end{figure}

Of particular interest is the time varying electric field $\mathcal{E}%
_{R}=\tfrac{Q}{R^{2}}$ on the external surface of the shell. In order to study
the variability of $\mathcal{E}_{R}$ with time it is useful to consider in the
tridimensional space of parameters $(R,T,\mathcal{E}_{R})$ the parametric
curve $\mathcal{C}:\left(  R=\lambda,\quad T=T(\lambda),\quad\mathcal{E}%
_{R}=\tfrac{Q}{\lambda^{2}}\right)  $. In astrophysical applications
\cite{RVX02} we are specially interested in the family of solutions such that
$\frac{dR}{dT}$ is 0 when $R=\infty$ which implies that $\Gamma=1$. In fig.
\ref{fig1} we plot the collapse curves in the plane $(T,R)$ for different
values of the parameter $\xi\equiv\frac{Q}{M}$, $0<\xi<1$. The initial data
$\left(  T_{0},R_{0}\right)  $ are chosen so that the integration constant in
equation (\ref{CASO2}) is equal to 0. In all the cases we can follow the
details of the approach to the horizon which is reached in an infinite
Schwarzschild time coordinate. In fig. \ref{fig2} we plot the parametric
curves $\mathcal{C}$ in the space $(R,T,\mathcal{E}_{R})$ for different values
of $\xi$. Again we can follow the exact asymptotic behavior of the curves
$\mathcal{C}$, $\mathcal{E}_{R}$ reaching the asymptotic value $\frac{Q}%
{r_{+}^{2}}$. The detailed knowledge of this asymptotic behavior is of great
relevance for the observational properties of the EMBH formation, see e.g.
\cite{RV02}, \cite{RVX02}.

\newpage

\newpage

\end{document}